\documentclass{llncs}
\usepackage{graphicx}
\usepackage{amsmath}
\usepackage{amssymb}
\usepackage{stmaryrd}
\usepackage{epic}
\usepackage{curves}
\usepackage{url}

\newcommand{\lfp}{{\sf lfp}}

\newcommand{\glb}{{\sf glb}}

\def\ll{[\![}
\def\rr{]\!]}

\newcommand{\sset}[2]{\left\{~#1  \left|
                               \begin{array}{l}#2\end{array}
                          \right.     \right\}}

\newcommand{\qin}{\hspace*{0.15in}}
\newcommand{\tqin}{\hspace*{0.45in}}
\newenvironment{SProg}
     {\begin{small}\begin{tt}\begin{tabular}[t]{l}}%
     {\end{tabular}\end{tt}\end{small}}
\def\anno#1{{\ooalign{\hfil\raise.07ex\hbox{\small{\rm #1}}\hfil%
        \crcr\mathhexbox20D}}}

\title{Experiments with a Convex Polyhedral Analysis Tool for Logic Programs}
\author{Kim S. Henriksen
\and Gourinath Banda
\and John P. Gallagher
}
\institute{Computer Science, Building 42.2, P.O. Box 260,\\
Roskilde University, DK-4000 Denmark\\
Email: {\tt \{kimsh,gnbanda,jpg\}@ruc.dk}
}
\begin{document}
\maketitle

\begin{abstract}
Convex polyhedral abstractions of logic programs have been found very useful in deriving numeric relationships between program arguments in order to prove program properties and in other areas such as termination and complexity analysis. We present a tool for constructing polyhedral analyses of (constraint) logic programs.  The aim of the tool is to make available, with a convenient interface, state-of-the-art techniques for polyhedral analysis such as delayed widening, narrowing, ``widening up-to", and enhanced automatic selection of widening points. The tool is accessible on the web, permits user programs to be uploaded and analysed, and is integrated with related program transformations such as size abstractions and query-answer transformation.  We then report some experiments using the tool, showing how it can be conveniently used to analyse transition systems arising from models of embedded systems, and an emulator for a PIC microcontroller which is used for example in wearable computing systems.  We discuss issues including scalability, tradeoffs of precision and computation time, and other program transformations that can enhance the results of analysis.
\end{abstract}

\section{Introduction}\label{sec:introduction}

Convex polyhedron analysis was first described by Cousot and Halbwachs 
\cite{Cousot-Halbwachs-78}. It is an application of abstract interpretation where
the program to be analysed is interpreted over the abstract domain of convex polyhedra. 

In logic programs, predicates whose variables range over numbers define numeric relations. Each $n$-ary numeric predicate is approximated by an $n$-dimensional polyhedron describing
linear relationships between its arguments.  For programs whose variables do not range over numbers, some abstraction such as a size norm is applied before computing a polyhedron approximation. 
In (constraint) logic programming (CLP), convex polyhedron
analysis has been used as part of termination analysers \cite{akinglowerbound}, 
for time-complexity analysis and argument-size analysis 
\cite{Benoy-King-LOPSTR96,benoy02computing}. 

Our interest lies both in the above application areas and also in analysing programs that model transition systems and operational semantics of various kinds, in order to discover invariant relations among state variables, or to check whether certain numeric relationships ever occur in reachable states.  We discuss such applications later in the paper. The use of polyhedra for analysis of hardware and software has of course been investigated by others (e.g. \cite{halbwachs97verification,BessonJT99,BagnaraHZ07TR}); our approach is to define mappings from such systems to CLP models, hence transforming the problem to a CLP analysis problem and allowing a single tool to analyse many formalisms.

In this paper we report on a tool, whose aim is to assist in exploring the effectiveness of convex polyhedron analysis.  
The convex polyhedron domain is an example of an infinite-height abstract domain, and such domains require some method of ensuring termination of the analysis. Typically widening
is used for this purpose. Widening introduces a trade-off between precision of the
approximation and efficiency of the analyser. Different methods for improving the
precision of widening for the convex polyhedral domain have been suggested
\cite{DBLP:conf/cav/GopanR06,Benoy-King-LOPSTR96,halbwachs97verification,WideningLandmarks}.  The tool allows various options to be selected easily, and the results displayed.  The tool also incorporates size abstractions and permits both goal-independent and goal-dependent analysis, the latter via query-answer transformation.  Since the tool is intended as an aid to experimentation, a {\em verbose} option allows a trace of the analysis to be viewed; this is especially useful in understanding where information is lost due to widening and the convex hull operation.

\paragraph{Outline of the Paper.}
Section \ref{sec:convexpolyhedra} will describe the convex polyhedron domain and 
operations required for an analyser for this abstract domain. Section 
\ref{sec:aiconvexpolyhedra} describes the abstract domain of convex polyhedra for
CLP programs. Section \ref{sec:tool}
describes the implementation details of our convex polyhedron analyser and
a web interface for the analyser.
Section \ref{sec:experiments} contains experimental results for the analyser.

\section{Convex Polyhedra}\label{sec:convexpolyhedra}
Polyhedra are geometric representations of linear systems of equalities and inequalities.
A convex polyhedron is a region of an $n$-dimensional space that is bounded by a
finite set of hyperplanes. For {\em not necessarily closed} (NNC) convex polyhedra 
strict inequalities are allowed in the representation of the polyhedron.
Certain constraint systems cannot easily be described using only closed
convex polyhedra. Accurate descriptions would require the use of open inequalities.

The set $\mathcal{P} \subseteq \mathbb{R}^{n}$ is a {\em not necessarily closed}
polyhedron (NNC polyhedron) if either
\begin{itemize}
\item[-] it is the intersection of a 
finite family of open or closed linear halfspaces
of the form $\{x | ax \ge c\}$ or $\{x | ax > c\}$ where 
$\mathbb{R}^{n}$ is the $n$-dimensional vector space
on real numbers, $a \in \mathbb{R}^{n}$ is a non-zero row vector and $c \in \mathbb{R}$ 
is a scalar constant  and $ x = \langle x_1, \ldots, x_n \rangle$
\item[-] $ n=0$ and $\mathcal{P} = \emptyset$
\end{itemize}
The set of all NNC polyhedra on the vector space $\mathbb{R}^{n}$ is denoted $\mathbb{P}_{n}$.
Each polyhedron is the set of solutions to a constraint system.
$$
\mathcal{P} \overset{def}{=} \{ x \in \mathbb{R}^{n} | A_{1}x = b_{1}, 
      A_{2}x \ge b_{2}, A_{3}x > b_{3}\},
$$
where $\forall i \in \{1,2,3\}$, $A_i \in \mathbb{R}^{m_i} \times \mathbb{R}^n$ 
and $b_i \in  \mathbb{R}^{m_i}$ and $m_1$, $m_2$, $m_3 \in \mathbb{N}$ are the numbers
of equalities, inequalities and strict inequalities respectively.

We use $\mathcal{C}$ to denote the constraint representation of 
the polyhedron $\mathcal{P}$:
$$ \mathcal{P} = con(\mathcal{C}) $$

%






\subsection{Operations on polyhedra}\label{sec:polyops}
For a convex polyhedron analysis of a program a few operations on polyhedra are required.

\subsubsection{Intersection}
The {\em intersection} of two polyhedra
$\mathcal{P}_1,\mathcal{P}_2 \in \mathbb{P}_{n}$ is denoted 
$\mathcal{P}_1 \cap \mathcal{P}_2$, and it is the largest NNC polyhedron that is included in
both $\mathcal{P}_1$ and $\mathcal{P}_2$.


\subsubsection{Convex Hull}
The {\em convex polyhedral hull} of two polyhedra
$\mathcal{P}_1,\mathcal{P}_2 \in \mathbb{P}_{n}$ is denoted 
$\mathcal{P}_1 \uplus \mathcal{P}_2$, and it is the smallest NNC convex polyhedron that contains
both $\mathcal{P}_1$ and $\mathcal{P}_2$.
The convex hull is an upper approximation of union, since convex polyhedra generally
are not closed under union. 

\subsubsection{Emptiness}
Given a set of constraints the polyhedral representation of this set 
may be empty if the constraints are not satisfiable. 
Checking satisfiability of a set of {\em linear} constraints 
is decidable.

\subsubsection{Inclusion}
The convex polyhedron analysis proceeds as a fix point computation. Some
mechanism of establishing when convergence is reached is needed. This can be
achieved by checking inclusion between two polyhedra.  E.g.\ for all
$\mathcal{P}_1,\mathcal{P}_2 \in \mathbb{P}_{n}$, $\mathcal{P}_1 = con(\mathcal{C}_1)$ 
entails $\mathcal{P}_2 = con(\mathcal{C}_2$) iff $\mathcal{C}_1 \subseteq \mathcal{C}_2$.
This is also decidable.

\subsubsection{Projection}
The projection operation returns the most 
precise polyhedron which does not depend on a given dimension.
Given an $n$-dimensional polyhedron $\mathcal{P}$, the projection
$\mathcal{P}' = proj(\mathcal{P},j)$, will return the most precise polyhedron, $\mathcal{P}'$,  
of dimensions $n-1$ that is entailed by $\mathcal{P}$ excluding its constraints
on dimension $j$.

\section{Convex Polyhedral Domain for Abstract Interpretation}\label{sec:aiconvexpolyhedra}
This section describes how convex polyhedra can serve as an abstract domain
for abstract interpretation of logic programs.

\subsection{Partial Ordering on Polyhedra}
Polyhedra can be ordered in a partially ordered set 
 where inclusion can be used as the ordering  
($\langle \mathbb{P}_n, \subseteq \rangle$). We can add
intersection as the greatest lower bound ($\bigsqcap$), the convex polyhedral
hull as the least  upper bound ($\bigsqcup$) and $\mathbb{R}^n$ as top element 
and $\emptyset$
as the bottom element; thus we have a lattice 
$\langle \mathbb{P}_{n}, \subseteq, \cap, \uplus, \emptyset, \mathbb{R}^n \rangle$.
This property makes convex polyhedra a suitable domain for program analysis - in
particular for abstract interpretation. 
The lattice is not a complete lattice but completeness is not required for abstract
interpretation to be applied.


The polyhedron will provide an abstraction of each program point in the program, where
the definition of a program point would depend on both the programming language and 
the program analysis.
For a logic program the program points would be the predicates in that program.

The abstract domain is the set of mappings $(Pred \mapsto \mathbb{P})$ where
an $n$-ary predicate $p \in Pred$ is mapped to an $n$-dimensional 
polyhedron $\mathcal{P} \in \mathbb{P}_n$.
Inclusion is used as the ordering over the mappings. For instance, for the mappings 
$m_1$ and $m_2$ the ordering is
$$
m_1 \sqsubseteq m_2 \equiv \forall p \in Pred, ~ m_1(p) \subseteq m_2(p)
$$
We can also represent such a mapping as a set of constrained atoms
$$
\begin{array}{rl}
 &      p(x_1,\ldots,x_n) \leftarrow c(x_1,\ldots,x_n) \\
\text{where}  & \mathcal{P} = con(c(x_1,\ldots,x_n)) \text{ for some } n\text{-dimensional polyhedron } \mathcal{P} \\

\end{array}
$$
For a finite set of predicates $\{p_1,\ldots,p_k\}$ we can represent the
mappings as a tuple of polyhedra $\langle \mathcal{P}_1, \ldots, \mathcal{P}_k \rangle$.
The concretisation function of the mapping $m$ would be defined as 
$$
\gamma(m) = \{p(t_1,\ldots,t_n) | \langle t_1,\ldots,t_n \rangle \in m(p) \}
$$

The abstract domain for a program would typically be a tuple 
of polyhedra, one element for each 
program point. The lattice structure is an extension of the one defined above, for example 
$\langle P_1, P_2, \ldots, P_k \rangle \sqsubseteq \langle Q_1, Q_2, \ldots, Q_k \rangle$
iff $P_1 \subseteq Q_1 \land P_2 \subseteq Q_2 \land \ldots \land P_k \subseteq Q_k$.

\subsection{Widening for Convex Polyhedra}\label{sec:bhrz03}
In the domain of convex polyhedra infinite ascending chains can occur, hence
the ascending chain condition is not satisfied. For program
analysis purposes some mechanism for accelerating the
fix point computations to convergence
may be required to ensure termination. The most used mechanism for this
is widening \cite{Cousot-Cousot-92}.
Widening in the convex
polyhedra domain was defined by Cousot and Halbwachs \cite{Cousot-Halbwachs-78}
and refined in Halbwach's PhD thesis.
This widening operator is generally referred to as the {\em standard widening} and
few attempts have been made to improve the operator itself. Recently Bagnara {\em et al.} \cite{BagnaraPrecise}
suggested a framework for constructing improved widening operators for the 
convex polyhedral domain. Where the
standard widening operator is restricted to only looking at the constraint representation of a
polyhedron, the new operators can be based on both the constraint and the parametric
representation. In short, the method allows operators that otherwise do not
meet the widening operator criteria to be used, while still ensuring termination. 
The resulting widening operator is not guaranteed to be more precise than
the standard widening but it is never {\em less} precise.

\subsection{Concrete and Abstract Semantics for CLP}\label{fixpoint}
This section will describe the semantics of the convex polyhedron analyser.
A concrete semantics and an abstract semantics will be defined. The concrete semantics
is described similar to the $T_P$ semantics \cite{EmdenKowalski}.
The concrete domain will be over the set of all formulas, $Atom \leftarrow Con$, where
$Con$ is a constraint system.
The immediate consequence operator is defined as:
$$
\begin{array}{l}
T^{\mathcal{C} }_{P} : 2^{Atom \leftarrow Con} \rightarrow 2^{Atom \leftarrow Con}  \\
\\
T^{\mathcal{C} }_{P}(I) = \sset{A \leftarrow \mathcal{C}}
{A \leftarrow B_1,\ldots,B_n \in P \\
\{\langle B_1 \leftarrow \mathcal{C}_1 \rangle, \ldots, \langle B_n \leftarrow \mathcal{C}_n \rangle \} \in I \\
\mathcal{C}' = \underset{i=1,\ldots,n}{\bigwedge} \mathcal{C}_i \\  
\mathcal{C}' \not\equiv false, \\

\mathcal{C} = \mathit{project}(\mathcal{C}', \mathit{Var}(A)) \\
}\\
\\
M^{\mathcal{C}} \ll P \rr = \lfp(T^{\mathcal{C} }_{P} ) 
\end{array}
$$
\noindent It is assumed that all built-in constraint predicates are in $I$ (e.g.\ {\tt X < Y :- X < Y} $\in I$).
We also assume some adequate satisfiability and projection algorithm for constraints 
appearing in the program exists.
The set of constrained atoms $\{\langle B_1 \leftarrow \mathcal{C}_1 \rangle, \ldots, \langle B_n \leftarrow \mathcal{C}_n \rangle \}$ is renamed apart.
The domain of the interpretation is the powerset of the set of facts of the form
$p(X_1, ...,X_n) \leftarrow \mathcal{C}$, where 
$p$ is a predicate in $P$ and $\mathcal{C}$ is a set of constraints over $X_1, ...,X_n$.

The abstract semantics is defined next.
Let $T^{\mathcal{C} }_{P}$ be the concrete semantics function, then the abstract semantics
of the program $P$ is the fix point defined as
$$
\begin{array}{l}
T^{\mathbb{P} }_{P} : 2^{Atom \leftarrow Con} \rightarrow 2^{Atom \leftarrow Con}  \\
T^{\mathbb{P}}_P (I) = I \uplus T^{\mathcal{C} }_{P}(I) \\
M^{\mathbb{P}} \ll P \rr = {\sf fp}(T^{\mathbb{P}}_{P} ) \\
\end{array}
$$
where the convex hull operator is extended to operate on constrained atoms.
Clause heads are standardised so variables in the heads are renamed 
consistently and for example constants occurring in heads
are replaced with a variable and this variable and and its constraint is added to the clause
body.
The domain of the abstract semantics is the powerset of the set of facts of the form
$p(X_1, ...,X_n) \leftarrow \mathcal{P}^n$,
where $p$ is a predicate in $P$ and $\mathcal{P}^n$ is a NNC convex polyhedron.

\subsection{Precision Improvements}
Applying widening in convex polyhedral domain introduces a loss of precision.
Different strategies have been suggested for minimising the loss of precision.

\subsubsection{Narrowing for Convex Polyhedra}
Narrowing in the convex polyhedra domain has so far been unexplored \cite{BagnaraPrecise}.
It has previously been suggested that narrowing for the convex polyhedral domain
could produce more precision \cite{Benoy-King-LOPSTR96}, but no implementation
of narrowing has been proposed or experimented with.

Strictly speaking a narrowing operator must ensure eventual stabilisation according to the second
requirement of its definition: An operator $\Delta : L \times L \rightarrow L$ is a narrowing operator if and only if
\begin{enumerate}
\item[-] $\forall l_1, l_2 \in L : l_1 \sqsubseteq l_2 \Rightarrow l_1 \sqsubseteq (l_2 \Delta l_1)
\sqsubseteq l_2$
\item[-] for all descending chains $(l_n)_n$ the sequence $(l^{\Delta}_{n})_n$
         eventually stabilises
\end{enumerate}
Quoting Cousot and Cousot in \cite{Cousot-Cousot-92}:
\begin{quote}
 A simple narrowing is obtained by limiting the length of the decreasing iteration
sequence to some $k \ge 1$ (experience shows that $k > 1$ often brings no significant
improvement).
\end{quote}

This would indicate that it might be sufficient to look at the first requirement for
a narrowing operator, and settle for a relatively low number of iterations, 
ignoring the convergence requirement. 

As also suggested in \cite{Benoy-King-LOPSTR96} we propose the use of the 
greatest lower bound ($\glb$) as a simple narrowing operator. 
For polyhedra
the $\glb$ is the intersection operation. The use of the intersection operator is safe, in the sense
that  
it yields a safe approximation of the least fix point but it does not guarantee convergence.

The $\glb$ operator, in this case intersection, is applied at each program point, $p$, 
to the polyhedron derived from the widened fix point computations, $\mathcal{P}_{p_{\nabla}}$, 
and the polyhedron derived from applying the semantic transfer function $f$ to the
program point $p$. If narrowing results in a more precise approximation for some
program point $p$, then the narrowing procedure must be reiterated to ensure that the
most precise approximation is used for narrowing of those program points that depends on $p$:

$$
\begin{array}{ll}
\mathcal{P}_{p_{\Delta}} = \mathcal{P}_{p_{\nabla}} \sqcap f(\mathcal{P}_{p_{\nabla}}) & 
        \text{if} ~ \mathcal{P}_{p_{\Delta}} = \emptyset \\  
\mathcal{P}_{p_{\Delta}} = \mathcal{P}_{p_{\nabla}} \sqcap f(\mathcal{P}_{p_{\Delta}}) &  \text{otherwise}  \\
\end{array}
$$

\subsubsection{Delayed Widening}
This technique is quite simple. The application of the widening operator is delayed
for a number of iterations. This will allow the analyser to build a set of more explicit constraints
to widen on \cite{DBLP:conf/cav/GopanR06} and produce more precise analysis 
results as shown in \cite{Benoy-King-LOPSTR96}. 
%

\subsubsection{Widening with thresholds/widening up-to}\label{sec:widenupto}
This technique was described for the interval domain in \cite{BlanchetCousotEtAl02-NJ} and
the convex polyhedral domain in \cite{halbwachs97verification}.
In the {\em interval domain} variables in a program are abstracted by a tuple $[a,b]$ where $a \le b$ 
indicating a lower and upper bound on the values that the abstracted variables can have during execution of the program.
 
The need for this
arose from the fact that narrowing for the interval domain would not recover bounds on loop variables 
if the exit condition for the loop contained disequalities 
A finite set of {\em threshold} values, $T$, 
including $-\infty$ and $+\infty$ are used to find better approximations of lower or upper bounds, than
simply $-\infty$ and $+\infty$. Widening with thresholds for the interval domain is defined as
$$
\begin{array}{rl}
\lbrack a,b]~ \nabla_T ~ [a',b'] = [a_l,b_h] ~\text{where} & a_l =
\begin{cases}
a & \text{if}~  a' \ge a \\
max\{l \in T | l \le a'\} & \text{otherwise}
\end{cases} \vspace{2mm}\\
\text{and} & b_h = 
\begin{cases}
 b & \text{if} ~ b' \le b \\
 min\{h \in T | h \ge b'\} & \text{otherwise}
\end{cases} \\
\end{array}
$$
The set of thresholds may be derived from the analysed program itself \cite{halbwachs97verification} or 
otherwise specified by the user.

We will implement this in the convex polyhedron analyser for constraint logic programs described later.
We suggest initially to derive the set of widening-up-to constraints, 
which we will call the {\em bounding convex polyhedron}, 
from the analysed program itself. For constraint logic programs, this set can easily be obtained for 
each predicate in the program, by taking the convex hull of the polyhedra derived from 
intersecting the constraints  on the built-in arithmetic predicates occurring in each clause body. 

\vbox{
\begin{definition}[Clause constraints]
The set of clause constraints for a program $P$ is defined as:
$$
\begin{array}{l}
\mathcal{C}(P) = \sset{A \leftarrow \mathcal{C}}
{ A \leftarrow B_1,\ldots,B_n \in P \\
        \mathcal{C} = \bigcap \sset{B_{lin}}
        {i \in \{1,\ldots,n\} \\
                B_{lin} = \mathit{linearise}(B_i)\\
        }
}
\end{array}
$$
where $\mathit{linearise}(B)$ returns the linear approximation of an atom $B$ if $B$ is an 
built-in arithmetic predicate, else the atom $B$ will be unconstrained. 
\end{definition}
}
Non-linear built-ins, for example the bit wise {\em or}-operator can be given a 
linear approximation; in this case the expression {\tt X is Y $\backslash /$ Z} can be approximated by
the constraint $X \le Y + Z$. These approximations may not be exact and e.g.\ the {\em or}-operator
may not occur frequently in CLP programs but it broadens the set of programs the analyser
can handle.\\
\vbox{
\begin{definition}[Bounding convex polyhedron]
The bounding convex polyhedra for a program $P$ is defined as:
$$
\begin{array}{l}
%
 \mathcal{P}^{\hat{\nabla}}_P = \sset{p \leftarrow \mathcal{P}}
 { 
  p \in \Pi_P\\
 \mathcal{P} =  \biguplus p \leftarrow C \in \mathcal{C}(P) \\}
 \end{array}
$$
\end{definition}
}

As long as this set of ``widen-up-to'' constraints remains static each time the widening up-to operator
is applied, convergence is assured.

\subsubsection{Selecting Widening Points}\label{sec:cutloop}
Another way of minimising the loss of precision introduced by widening is
to find a good set of widening points, $W$, as small as possible.
When $W$ is chosen such that every loop
in the dependency graph contains at least one element also in $W$, then any chaotic iteration
strategy \cite{Cousot-Automatic-Synthesis,Cousot-Monotone-Equations} 
over the system of semantic equations will terminate with a safe approximation of $\lfp(f)$ \cite{Bourdoncle93}.
Bourdoncle \cite{Bourdoncle93} and Cousot \cite{CousotSemanticFoundations} suggested methods 
for selecting the set of widening points. Both suggested a method based on detecting
feedback edges in the dependency graph \cite{BaaseGelder}. The predicate call graph of typical
logic programs would contain many small strongly connected components, typically arising from
predicates with direct recusion. For imperative programs and in particular programs allowing
goto-statements, the dependency graph would contain few large strongly connected components. 
Applying the feedback edge detection to such programs may result in a set of widening points much larger than
the optimal set. 

We have implemented a simple cut-loop algorithm that for some dependency graphs will lead to a smaller set of widening points than the feedback edge detection. \vspace{4mm}

\vbox{
\noindent {\bf Input:} Dependency graph $G = (N,E)$ \\
{\bf Output:} Wideningpoints $W \subseteq N$ \\
\\
{\bf Begin} \\
\qin $loops = \emptyset$ \\
\qin for $n \in N$ \\
\qin \qin traverse($n,[~]_{ancestors}$) \\
\qin $W = \emptyset$ \\
\qin while $loops \neq \emptyset$ \\
\qin \qin for $n \in N$ \\
\qin \qin \qin $loopCount[n] = \vert \{l|l\in loops, n \in l\} \vert$ \\
\qin \qin $candidates = \{n \in N | loopCount[n] = max_{j \in N}(loopCount[j])\}$ \\
\qin \qin select $wp \in candidates$ \\
\qin \qin $loops = loops \setminus \{l \in loops | wp \in l \}$ \\
\qin \qin $W = W \cup \{ wp \} $\\
{\bf End} \\
}
\begin{center}
{\it (algorithm continues on the next page)}
\end{center}
\vbox{
traverse($n,n_{ancestors}$) $\{$ \\
\qin if visited($n$,$G$) then \\
\qin \qin if $n \in n_{ancestors}$ then \\
\qin \qin \qin $loop =$ path from $n \in n_{ancestors}$ to head of $n_{ancestors}$ \\
\qin \qin \qin $loops = loops \cup loop $\\
\qin \qin endif \\
\qin else \\
\qin \qin markVisited($n,G$) \\
\qin \qin for $n_s \in  Succ(n,G)$ \\
\qin \qin \qin traverse($n_s,[n | n_{ancestors}]$)\\
\qin endif \\
$\}$ \\
}

\begin{figure}[htp]
\begin{center}
        \includegraphics*[scale=0.65,viewport=120 500 380 675]{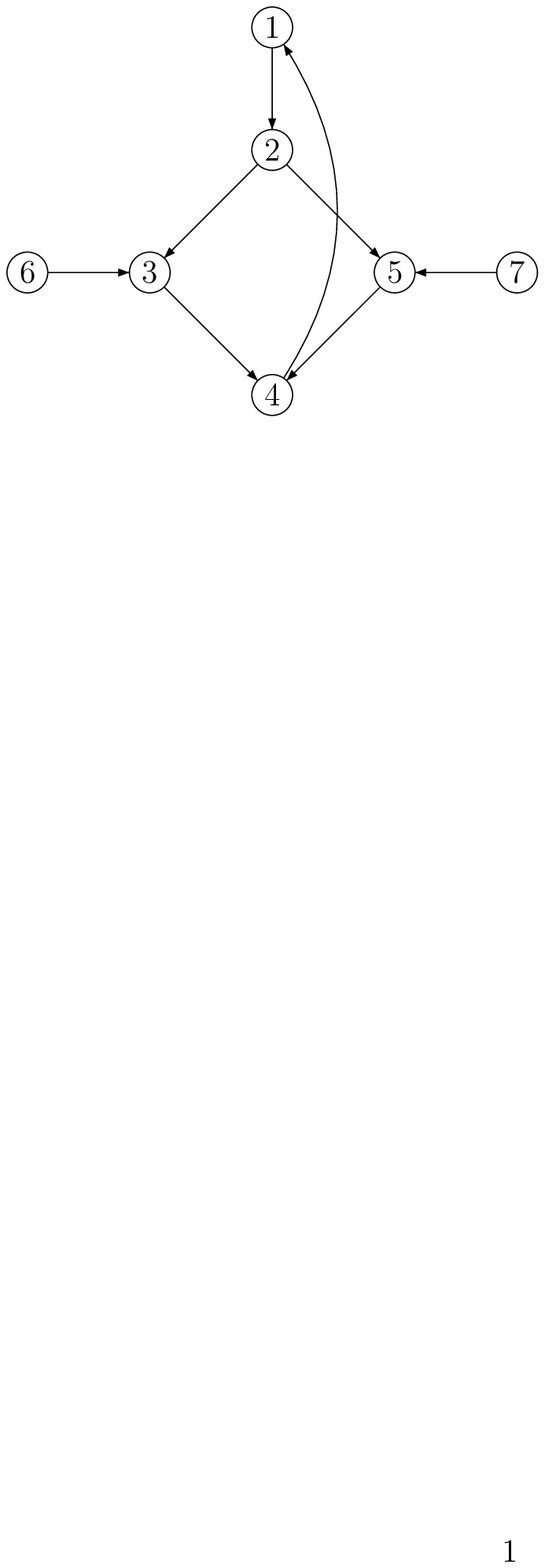}
\end{center}
\caption{Directed Graph with multiple entry nodes, e.g.\ $\{6,7\}$.}\label{fig:multi-entry-graph}
\end{figure}

The algorithm traverses the graph and records the loops found. The first widening point selected
is the node that is part of the highest number of loops. The set of loops that includes this widening
point is then eliminated. This procedure continues until all loops have been eliminated.
Figure \ref{fig:multi-entry-graph} shows a graph with a strongly connected component having
nodes that are part of more than one loop, these being the nodes 1, 2 and 4. Additionally
the graph has more than one entry node e.g.\ $\{6,7\}$. 
Applying the feedback edge detection to this graph
would result in the set of widening points being either $\{3,5\}$ or $\{4,5\}$ depending on
which entry node is chosen first. The cut-loop algorithm would choose only node 1 as widening point
for any node selected as the entry node.



\section{Description of the Analysis Tool}\label{sec:tool}

In this section we describe the analysis tool and the main features of its implementation. The polyhedral analysis algorithm itself, specified as the least fix point of the $T^{\mathcal{C} }_{P}$ operator in Section \ref{fixpoint} is implemented in Ciao Prolog \cite{ciao-reference-manual-tr}.  The initial implementation was 
based on a method for constructing bottom-up evaluator for logic programs, developed by Codish and S{\o}ndergaard \cite{Codish-jlp99,Codish-Sondergaard-02}. The naive bottom up interpreter is a small Prolog program, closely implementing the $T_P$ semantics. In each iteration, clause heads whose body can be proven from existing facts are asserted as new facts themselves. The program is evaluated iteratively until a fix point is reached. A fix point is reached, when no new heads can be asserted on some iteration.

For the convex polyhedron analyser the naive bottom up interpreter is modified so each fact is associated with a polyhedron.  Thus the predicate ``facts" have two arguments, an atom of the form $p(X_1,\ldots,X_n)$
and a set of constraints $\mathcal{C}$ over (some of) the variables $X_1,\ldots,X_n$. In addition we modified the naive fixpoint iteration to incorporate the well-known {\em semi-naive} optimisation and decomposition of the single fixpoint computation into a sequence of fixpoints for the strongly connected components of the predicate dependency graph.

\subsection{The Parma Polyhedra Library}\label{sec:ppl}
For handling the polyhedra, we use the Parma Polyhedra Library (PPL) (version 0.9), a programming library 
targeted especially at analysis and verification \cite{bagnara02possibly,BagnaraTowards}.
The PPL was chosen since it implements the operations needed for a convex polyhedron analyser and it has interfaces
for a variety of programming languages including Ciao Prolog \cite{ciao-reference-manual-tr}. It is also portable, free, well documented and well engineered.
In addition it supports operations on not necessarily closed polyhedra 
\cite{bagnara03new,bagnara02possibly}. 
Development began in 2001 and the library is under further development.
Other libraries for manipulating polyhedra exist, such as the NewPolka
library\footnote{http://pop-art.inrialpes.fr/people/bjeannet/newpolka/index.html}
and PolyLib\footnote{http://icps.u-strasbg.fr/PolyLib/}. 

The PPL interface to Ciao Prolog provides ``handles" to polyhedra, which are constants within Prolog code.  When passed as arguments to PPL functions they become keys giving access to specific polyhedra.  The predicate facts described above thus have PPL handles as their second arguments instead of sets of constraints. During the analysis the Prolog code does not manipulate constraints directly at all, apart from those appearing directly in the code of the program to be analysed. 

\subsubsection{Web interface}

The web interface has the following main functions.
\begin{itemize}
\item
Browsing and uploading of program from the user's machine.  Some prepared examples can also be selected.
\item
Abstraction of the uploaded program using size-norms (currently term-size and list-length norms are available).
\item
Selection of options for widening, delayed widening, narrowing (number of narrowing iterations), and widening up-to (two variants).
\item
Optional provision of widening points.
\item
Selection of output information, including a trace of the successive operations during the fixpoint computation and narrowing phase, timing information and constraint counts (a rough measure of precision).
\item Query-answer transformation of the uploaded program. This allows for analysis with
respect to a supplied goal.
\end{itemize}

\begin{figure}
\begin{center}
\includegraphics[width=\textwidth]{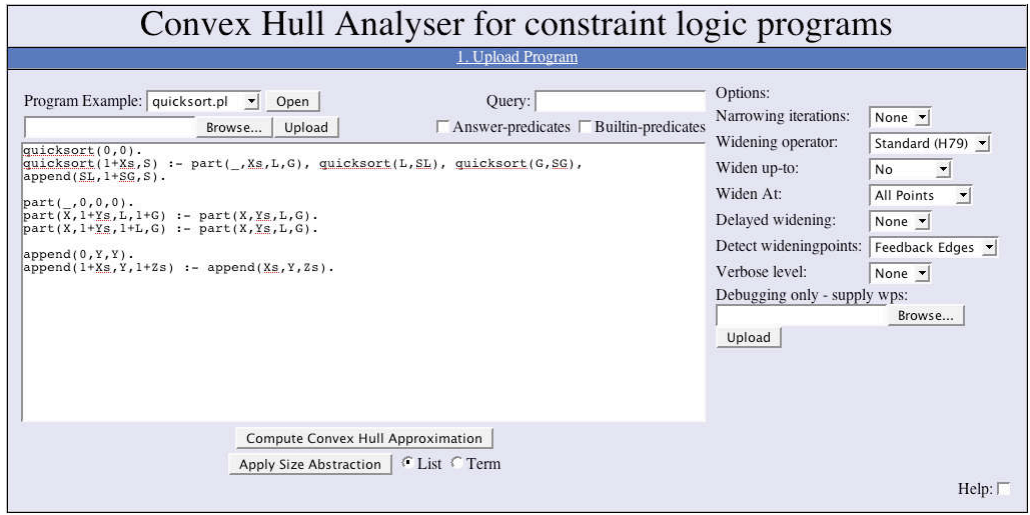}
\end{center}
\caption{Screenshot of web interface}\label{fig:screenshot1}
\end{figure}

The interface is implemented in PHP and calls the analyser using the 
script facility of Ciao Prolog, allowing a program to be executed from a command line. A
screenshot of the web interface is shown in Figure \ref{fig:screenshot1}.

The web interface is available at
\url{http://wagner.ruc.dk/CHA/}

\section{Experiments}\label{sec:experiments}
We describe some ongoing work in analysing CLP models of embedded systems and of a small microcontroller.

\subsection{Experiments with Embedded System Designs}\label{sec:lhaexperiments}
We have experimented with the CHA tool in analysing embedded systems that are (i) formally specified as Linear Hybrid Automata (LHA) \cite{alur93automatic}  or (ii) programmed in SIGNAL \cite{BenvenisteGJ91}. LHA is a formal specification language while SIGNAL is a programming language for reactive systems. Both these languages are extensively used in the embedded systems domain, where the systems are reactive and state transition systems.

In ongoing work, systematic semantic mappings are defined for each of these languages to transform the systems in these languages into CLP programs, which are analysed by the CHA tool to extract the linear invariants on the system variables. (Note that the CLP programs here are not necessarily intended to be executed, though they could be used to simulate behaviour in some cases.) We sketch in this section a water-level controller specified by an LHA model, taken from \cite{halbwachs97verification}, and a similar example of a tub-controller specified by a SIGNAL program \cite{BessonJT99}. We also proved properties of an LHA model of the Fischer mutual exclusion protocol, also taken from  \cite{halbwachs97verification}.

Rather than developing a customized solution algorithm and tool for each formalism,  our approach provides one single tool which could be employed for extracting linear invariants from any formal specification of a state transition system. Further, much of mathematics concerned with fixed-point equations is handled in a uniform way within the model semantics of CLP; all the user needs to provide is a CLP equivalent of the system to be analysed.  Of course, a correct translation of each formalism to CLP has to be performed, but we believe that this is generally a simpler task than writing a new analyser for each formalism.

\subsubsection{LHA.}

\begin{figure}[t]
\begin{center}
\includegraphics[width=2.5 in]{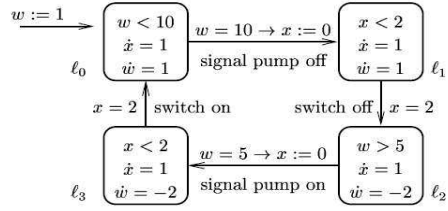}
\end{center}
\caption{A Water-level Monitor}
\protect\label{waterlevel}
\end{figure}

Informally an LHA is a state transition system, with locations and transitions. 
In locations the variables vary following a linear law respecting an imposed linear invariant on such variables, while transitions are triggered by the linear guards. 
The LHA model (quoted from \cite{halbwachs97verification}) for the water-level-controller is depicted in Figure \ref{waterlevel}. 
Boxes represent locations while arrows represent transitions. 
Each box (location) may contain one or more linear conditions over the variables, and constant varying rates of these variables with respect to time. 
The linear conditions are the invariants imposed on the corresponding locations. 
Such a system translates to the following semantically equivalent CLP program.

\begin{scriptsize}\begin{verbatim}

l0(W1,X1) :- W1=1.                           l0(W1,X1) :- l0(W,X), W1=W+1, W1<10, X1=X+1.
l1(W1,X1) :- l0(W,_), W=10, W1=W, X1=0.      l1(W1,X1) :- l1(W,X), W1=W+1, X1=X+1, X1<2.
l2(W1,X1) :- l1(W,X), X=2, W1=W, X1=X.       l2(W1,X1) :- l2(W,X), W1=W-2, W1>5, X1=X+1.
l3(W1,X1) :- l2(W,X), W=5, W1=W, X1=0.       l3(W1,X1) :- l3(W,X), W1=W-2, X1=X+1, X1<2.
l0(W1,X1) :- l3(W,X), X=2, W1=W, X1=X.
\end{verbatim}\end{scriptsize}

Though the above CLP program specifies the system behaviour faithfully, one can only observe the state variables at discrete time points as we model time growing with a granularity of 1 time unit. Therefore our analysis is based on this discrete state space. 

The range of the water level is given by the first argument in the predicates {\tt l0} to {\tt l3}. By adding extra clauses we can easily project out the constraints on this argument.
\begin{scriptsize}\begin{verbatim}
w0(W) :- l0(W,_).  w1(W) :- l1(W,_).  w2(W) :- l2(W,_).  w3(W) :- l3(W,_).
\end{verbatim}\end{scriptsize}
Computing the model of the above program with these extra projection clauses, we obtain the bounds on the water level in each state.
\begin{scriptsize}\begin{verbatim}
w0(A) :-  -1*A> -10,1*A>=1.    w1(A) :- -1*A> -12,1*A>=10.
w2(A) :- -1*A>= -12,1*A>5.     w3(A) :- -1*A>= -5,1*A>1.
\end{verbatim}\end{scriptsize}
These are the same results reported in \cite{halbwachs97verification}, where the reachable locations are characterized by a fixed point equation on the forward collecting semantics of LHA, and the abstract interpretation is applied to automatically compute the upper approximation of the solution to this fixed point equation. 

We also succeeded in verifying a property of the Fischer mutual exclusion protocol, also taken from \cite{halbwachs97verification}.  In this case, the task is to show under which conditions a critical section could be entered by more than one process simultaneously.  Space does not permit the detailed description of the system and its translation to CLP (please refer to the cited paper). The generated constraints on the illegal state {\tt l5} were the following (again, identical to the constraint derived in \cite{halbwachs97verification}).
\begin{scriptsize}\begin{verbatim}
l5(A,B,C,D) :- 11*A+ -10*B+10*C+ -11*D>=0,11*A+ -10*B>=0,1*A+ -1*D>=0,10*C+ -9*D>=0,1*D>=0,-9*A+10*B>=0.
\end{verbatim}\end{scriptsize}
From this it can be deduced that the state is unreachable if {\tt 9*D > 10*C}.  The unreachability of this state can also be verified in the CHA tool by adding this constraint to the initial state. The empty polyhedron is then obtained as the solution to predicate {\tt l5}.

\subsubsection {SIGNAL.}
SIGNAL \cite{BenvenisteGJ91} is a programming language for reactive systems, which is very popular in implementing mission critical systems. Again the SIGNAL program is systematically translated into a constraint logic program, and is subjected to convex polyhedron analysis by CHA tool. The following is the specification of a tub controller (quoted from \cite{BessonJT99}) with a tap (faucet) and a pump in SIGNAL.

\begin{scriptsize}\begin{verbatim}
( level := zlevel + faucet - pump
| faucet := zfaucet + ((1 when zlevel <= 4) default (-1 when zfaucet > 0)
default 0)
| pump := zpump + ((1 when zlevel >= 7) default (-1 when zpump > 0)
default 0)
| alarm := (0 >= level) or (level >= 9)
5
)
init zlevel = 1; zfaucet = 0; zpump = 0; zalarm = false
\end{verbatim}\end{scriptsize}

In the given SIGNAL program, sub-systems $level$, $faucet$, $pump$  and $alarm$ are composed in parallel, where $level$ computes the water-level; $faucet$ computes the opening position for the faucet; $pump$ computes the pumping rate of water; and $alarm$ raises an alarm when the water-levels exceed the set limits. A CLP translation, modelling the states in the associated transition semantics of SIGNAL, is as follows.

\begin{scriptsize}\begin{verbatim}
tubsystem :-                               levelLogic(L,P,T,L1) :- L1 is L+T-P.
      tubsystemStates(0,0,1).              alarmLogic(L,0) :- L > 0, L < 9.
tubsystemStates(A,B,C) :-                  alarmLogic(L,1) :- L =< 0.      
      faucetlogic(C,A,F),                  alarmLogic(L,1) :- L >= 9.
      pumplogic(C,B,G),                    faucetlogic(L,T,T1) :- L =< 4, T1 is T+1.
      levelLogic(C,G,F,H),                 faucetlogic(L,T,T1) :- L > 4, T > 0, T1 is T-1.
      alarmLogic(H,I),                     faucetlogic(L,T,0) :- L > 4, T =< 0.
      tubsystemStates(F,G,H).              pumplogic(L,P,P1) :- L >= 7, P1 is P+1.
tubsystemStates(A,B,C).                    pumplogic(L,P,P1) :- L < 7, P > 0, P1 is P-1.
                                           pumplogic(L,P,0) :- L < 7, P =< 0.
\end{verbatim}\end{scriptsize}
Note that in order to analyse the reachable states we need to analyse the calls to the predicate {\tt tubsystemStates} starting from the initial call {\tt tubsystem}, which is made possible in the CHA tool by selecting the query answer transformation option.  However such an initial attempt to prove the invariants was unsuccessful. The constraints were over general and indicated that the alarm could possibly be activated (which in fact it cannot). Also the attempts to overcome the problem by transforming the above program, creating multiple versions of the predicate {\tt tubsystemStates} for various cases did not solve the problem. The property that the alarm does not go off can be verified by first specialising the program with respect to the initial goal;  the use of specialisation for model checking was described by Leuschel {\it et al.} \cite{LeuschelG01}. Though the solution in this case is ad hoc, the fact that standard equivalence preserving CLP transformations can be applied to the program provides, we argue, further grounds for believing that CLP modelling and analysis of systems is a flexible and powerful approach.  Further work will aim at a more systematic integration of polyhedral analyses and program transformation.

\subsection{Experiments with a model of the PIC microcontroller}
The convex polyhedron analyser has been used to analyse CLP programs derived from
specialisation of a CLP based emulator for the PIC microcontroller \cite{KSH-JPG-SCAM}.
These programs are fairly large as Table \ref{tab:scalability} shows. 
Each instruction in the PIC program will generate an equivalent predicate in the specialised
CLP program. A typical predicate would look as shown:

\begin{SProg}
execute\_\_6(A,B,C,D) :- \\
\tqin      E is A+C,   F is E>>8, \\ 
\tqin      E$\backslash$==0,  F$\backslash$==0, \\
\tqin      G is 24$\backslash$/1,  H is D+1, \\
\tqin      execute\_\_7(E,B,C,H). \\
\end{SProg}

Each argument of the execute-predicates contains an element of the live machine state
at that program point. These elements are accumulator, clock, data registers and stack.
The analysis of the program containing the predicate shown above results in the
following constraints on the arguments: {\tt 
1*D>=5,4*B+1*D=45,2*A+ -5*D= -25,1*C=10}.

\subsubsection{Scalability}
The three test
case programs have from 200 to 600 clauses. The predicates of the largest program 
have on average 18 arguments. For these test results no improved widening techniques
were enabled. The tests shows that even larger CLP programs can be analysed in just one
minute. Timing results were collected on a machine running Linux equipped with a 1GHz 
Pentium III processor and 256MB RAM.

\begin{table}[hbp]
\begin{center}
\begin{tabular}{|l|c|c|c|c|c|}
\hline
{\bf Program} & {\bf No. Clauses} & {\bf Size (kb)} & {\bf Avg. arity} & 
\begin{tabular}{c}
{\bf Iterations} \\
{\bf to fixpoint} \\
\end{tabular} 
&
\begin{tabular}{c}
{\bf Analysis} \\
{\bf Time (sec.)} \\
\end{tabular} \\
\hline
Compass & 199 & 16 & 3.5 & 158 & 6 \\
Accelerometer & 274 & 19 & 3.2 & 127 & 26 \\
GPS & 631 & 94 & 18 & 209 & 70 \\
\hline
\end{tabular}
\end{center}
\caption{Analysis time for specialised CLP programs}\label{tab:scalability}
\end{table} 


\subsubsection{Precision}
Analysing larger programs such as the ones shown in Table \ref{tab:scalability}
produces sets of constraints too large to show here. Generally speaking a set 
containing a higher
number of constraints also represent a more precise approximation, so we use the number of constraints as a crude measure of precision. The results shown
below therefore only lists the total number of constraints for the whole program.
Different combinations of widening techniques have been applied to the same
program. The table shows that for the compass program the most precise
result is obtained combining all the improved widening techniques. For the
accelerometer program narrowing and delayed widening gives no increased precision in 
combination with widening up-to.

 \begin{table}[hbp]
\begin{center}
\begin{tabular}{|l|c|c|c|c|}
\hline
{\bf Program} & 
\begin{tabular}{c}
{\bf Delayed} \\
{\bf Widening} \\
{\bf {\small (iterations)}} \\
\end{tabular} 
& 
\begin{tabular}{c}
{\bf Simple } \\
{\bf Narrowing} \\
{\bf {\small (iterations)}} \\
\end{tabular} 
 & 
\begin{tabular}{c}
{\bf Widen } \\
{\bf up-to} \\
\end{tabular} 
 & 
 \begin{tabular}{c}
{\bf Resulting} \\
{\bf Constraints} \\
{\bf {\small (Number of)}} \\
\end{tabular}  \\
\hline
Compass &  &  & & 200 \\ 
Compass & & 200 & & 209 \\ 
Compass &  &  & $\surd$ & 302 \\ 
Compass &10  &  & & 354 \\ 
Compass & 10 & 200 & & 357 \\ 
Compass & 10 &  & $\surd$ & 362 \\ 
Compass & 10 &  200 & $\surd$ & 363  \\ 
\hline
Accel. & & & & 420 \\ 
Accel. & & 300 & & 420 \\ 
Accel. & 10 & & & 420 \\ 
Accel. &  & & $\surd$& 426 \\ 
Accel. &  10 & 300 & $\surd$& 426 \\ 
\hline
\end{tabular}
\end{center}
\caption{Improved Widening Techniques}\label{tab:improvedwidening}
\end{table}

\subsubsection{Selection of widening points}
Precision and efficiency of the analysis can also be affected by the choice of widening points; generally, the fewer widening points the better.
The specialised PIC programs have a dependency graph similar to that of imperative 
programs (few large strongly connected components) rather than that of a more
typical logic program (many small strongly connected components). 
Table \ref{tab:widenevalfeedbackresults} shows the compass program analysed
using both the feedback edge detection of widening points and the cut-loop algorithm
outlined in Section \ref{sec:cutloop}. Analysing using the smaller set of widening points
can produce more precise results, though for this example no better approximation can
be achieved than using all the improved widening techniques.

 \begin{table}[hbp]
\begin{center}
\begin{tabular}{|c|c|c|c|c|}
\hline
\begin{tabular}{c}
{\bf {\small Delay}} \\
{\bf {\small Widen}} \\
\end{tabular} 
& 
\begin{tabular}{c}
{\bf {\small Narrow}} \\
\end{tabular} 
 & 
\begin{tabular}{c}
{\bf {\small Widen }} \\
{\bf {\small up-to}} \\
\end{tabular} 
 & 
 \begin{tabular}{c}
{\bf {\small Feedback-edge}} \\
{\bf {\small  22 $\nabla$-points}} \\
\end{tabular} 
 & 
 \begin{tabular}{c}
{\bf {\small Cut-Loop}} \\
{\bf {\small 12 $\nabla$-points}} \\
\end{tabular}  \\
\hline
  &  & & 200 & 200 \\ 
 & 200 & & 209 & 209 \\ 
  &  & $\surd$ & 295 & 302 \\ 
10  &  & & 340  &  354\\  
 10 &  & $\surd$ & 355 &357  \\ 
 10 & 200 & & 357 &  362\\ 
 10 &  200 & $\surd$ & 363 & 363 \\  
\hline
\end{tabular}
\end{center}
\caption{Feedback-edge results compared to Cut-Loop results}\label{tab:widenevalfeedbackresults}
\end{table}

\subsection{Comparison with existing convex polyhedron analyser}
A convex hull analyser for CLP programs was reported in 
\cite{Benoy-King-LOPSTR96,BenoyPolyhedral}\footnote{This analyser is also available online
through the Ecce specialiser \cite{LeElVaCrFo06_140} \\ \url{http://www.stups.uni-duesseldorf.de/~asap/asap-online-demo/mecce.php}}. 
This analyser was compared to
previous analysers selecting the tricky test cases for comparison. Applying our
analyser to the same set of test case programs we obtain equally precise approximations.
Compared to the Benoy and King analyser ours has been extended with a few 
improved widening techniques and program transformation tools. This allows a wider
range of programs to be analysed using our tool. For instance the following program
can be analysed with respect to the query $exp(\_,10,\_)$ and with a single narrowing
operation applied it provides both an upper and lower bound on the second argument
of $exp\_/4$. \vspace{2mm}

\begin{SProg}
exp(X,Y,Z) :- exp\_(X,Y,1,Z). \\
 \\
exp\_(\_X,0,Ac,Ac). \\
exp\_(X,Y,Ac,Z) :- \\
\tqin	Y > 0, \\
\tqin	NewAc is X*Ac, \\
\tqin	NewY is Y-1, \\
\tqin	exp\_(X,NewY,NewAc,Z). \\
\end{SProg}

\section{Conclusion}\label{sec:conclusion}
We have developed a convex polyhedron analyser for constraint logic programs. The
analysis tool has been integrated with program transformation techniques 
including size abstractions and query-answer transformations. The analyser has been extended with some improved widening and narrowing techniques. The tool has been made available online for convenient experimentation. 

We have applied the analyser to variety of CLP programs, including CLP programs  automatically
derived from partial evaluation and embedded systems modelled in CLP.  This is ongoing work but the tool has demonstrated its worth in providing the means for experimenting with different ways of gaining precision for each case study, and helping with understanding (via the ``verbose" output) the reasons for imprecision when this occurs.

\subsection{Future Work}
A few widening strategies aiming at improving the precision of convex polyhedral analysis
exist that are not part of our analyser tool. 
Widening with landmarks \cite{WideningLandmarks}
shares some common traits with 
widening with thresholds. Where upper or lower bounds would be lost using standard
widening, they can in some cases be recovered using narrowing. The widening with
landmarks is a refinement of widening with thresholds, that will produce results
precise enough that narrowing would not be needed to recover lost bounds.
The look-ahead widening \cite{DBLP:conf/cav/GopanR06} is a recent method. 
Two polyhedra are used for abstracting each program point; a main
and a pilot polyhedron. Widening and subsequent narrowing are only performed on the pilot polyhedron.
The program is evaluated with respect to the main polyhedron, and program points that are not reachable
under the main polyhedron are ignored. Once the pilot stabilises it is promoted to a main polyhedron and
the program is reevaluated. This technique tries to solve the situations where during a loop a variable
may be either increasing or decreasing. Widening in this situation may cause a loss of both upper
and lower bounds. 
Both widening methods could be included in our tool.


Automatic program transformations to improve precision are also of great interest in future work.  In particular, multiple specialisations of one kind or another can preserve separate polyhedral approximations which would otherwise be merged by the convex hull operation, losing precision.  To illustrate this principle, consider McCarthy's 91-function.
\begin{scriptsize}\begin{verbatim}
mc91(N,X) :-  N > 100,  X is N - 10.
mc91(N,X) :- N =< 100,  Y is N + 11, mc91(Y,Y2), mc91(Y2,X).
main(X,N) :- X =< 100, mc91(X,N).
\end{verbatim}\end{scriptsize}
Here we provide a {\tt main} call where the argument is at most 100, and the result in this case is 91 (if the input is an integer). The analyser is not able to deduce this, returning only a result that the answer is greater than 90.  If we make two versions of the {\tt mc91} predicate, one ({\tt mc91h}) handling arguments greater than 100 and the other ({\tt mc91l}) handling arguments at most 100, we obtain the following program.
\begin{scriptsize}\begin{verbatim}
mc91l(N,Out):- N=<100, mc91h(N+11,X), mc91h(X,Out).
mc91l(N,Out):- N=<100, mc91h(N+11,X), mc91l(X,Out).
mc91l(N,Out):- N=<100, mc91l(N+11,X), mc91h(X,Out).
mc91l(N,Out):- N=<100, mc91l(N+11,X), mc91l(X,Out).
mc91h(N,X) :-  N > 100, X is N - 10.

main(X,Y):- X=<100, mc91l(X,Y).
\end{verbatim}\end{scriptsize}
Analysing this program with the query {\tt main(X,Y)} we obtain the result {\tt main\_ans(A,B) :- A =< 100, B > 90, B =< 91.} This is the most precise possible result and indeed there is only one integer, namely 91, in the solution for {\tt B}. (Note that the predicate {\tt main\_ans} is generated by the query-answer transformation).

Such transformations need to be studied more thoroughly;  in this case a simple enumeration of the possible calls to the two versions suffices, but this is a naive approach in general and would cause the program size to blow up.  A semantics-based approach based on Winsborough's multiple specialisation technique \cite{Winsborough92} would work in some case, though not in the one above.  A ``bottom-up" version of the multiple specialisation analysis 
is a possible future investigation.

\bibliographystyle{abbrv}
\bibliography{refs,gnbanda}

\end{document}